\documentclass[preprint,review,12pt]{revtex4}
\usepackage{amsmath}
\usepackage{color}
\usepackage[dvips]{graphicx}
\begin{document}
\title{Improved Equation of State for Finite-Temperature Spin-Polarized Electron Liquids on the Basis of Singwi-Tosi-Land-Sj\"{o}lander Approximation}

\author{Shigenori Tanaka}
\email{tanaka2@kobe-u.ac.jp}
\affiliation{Graduate School of System Informatics, Kobe University, 1-1 Rokkodai, Nada, Kobe 657-8501, Japan}

\date{\today} 

\begin{abstract}

An accurate expression for the exchange-correlation free energy $f_{xc}$ of homogeneous electron fluids at finite temperatures is presented 
on the basis of Singwi-Tosi-Land-Sj\"{o}lander (STLS) approximation. 
In addition to the derivation for the paramagnetic state, that for the ferromagnetic state is newly carried out in which 
a correction in the strong Coulomb coupling regime is incorporated into the construction of analytic expression as a function of the coupling constant and the Fermi degeneracy.
The fitting formula for $f_{xc}$ is then extended over any degree of spin polarization with the aid of available interpolation scheme.
The proposed equation of state, called iSTLS formula, shows reasonable agreements with the existing quantum Monte Carlo evaluations at finite temperatures 
in the paramagnetic state, thus giving a consensus for the thermodynamic functions between many-body theories and computer simulations.
On the other hand, the current status for the agreement among various evaluations of $f_{xc}$ is relatively unsatisfactory in the ferromagnetic state, 
suggesting the necessity of further investigations.
\\



\end{abstract}

\maketitle

\setlength{\baselineskip}{24pt}

\section{Introduction}

This paper considers a homogeneous electron fluid specified by the average number density $n = N/V$ and the absolute temperature $T$, where 
$N$ and $V$ are the particle number and the volume, respectively, with neutralizing positive-charge background. 
The Coulomb coupling constant for the degenerate electron gas or liquid is given by \cite{Ichimaru1,Ichimaru2} 

\begin{equation}
r_{s} = a/a_{B},
\end{equation}

\noindent
where $a = (3/4\pi n)^{1/3}$ is the Wigner-Seitz radius and $a_{B} = \hbar^{2}/me^{2}$ is the Bohr radius 
with $\hbar$, $m$ and $-e$ being the Planck constant, the mass and electric charge of electron, respectively.
In the high-temperature, classical limit, the strength of Coulomb coupling can instead be measured by \cite{Ichimaru1,Ichimaru2}

\begin{equation}
\Gamma = \frac{e^{2}}{ak_{B}T},
\end{equation}

\noindent
where $k_{B}$ refers to the Boltzmann constant.
The degree of Fermi degeneracy is then measured by \cite{Ichimaru1,Ichimaru2}

\begin{equation}
\theta = k_{B}T/E_{F},
\end{equation}

\noindent
where $E_{F} = \hbar^{2}k_{F}^{2}/2m$ is the Fermi energy with $k_{F} = (3\pi^{2}n)^{1/3}$ being the Fermi wavenumber. 
It is noted that the common $E_{F}$ with $k_{F}$ (not with $k_{F}' = (6\pi^{2}n)^{1/3}$) is employed both for the paramagnetic and ferromagnetic states of electron fluid 
in the present study, and thus $\theta$ is determined only in terms of $T$ and $n$.
There is a useful relation among the three parameters, $r_{s}$, $\Gamma$ and $\theta$, as 

\begin{equation}
\Gamma\theta = 2\lambda^{2}r_{s}
\end{equation}

\noindent
with $\lambda = (4/9\pi)^{1/3}$.
Moreover, the degree of spin polarization in the electron fluid is measured by \cite{Tanaka3}

\begin{equation}
\zeta = (n_{1}-n_{2})/n,
\end{equation}

\noindent
where $n_{1}$ and $n_{2}$ represent the number densities of spin-up and spin-down electrons, respectively, hence 
$\zeta = 0$ for the paramagnetic (spin unpolarized) state and $\zeta = 1$ for the ferromagnetic (spin polarized) state.

\par

Recently, there has been an increasing interest in the thermodynamic properties of uniform electron gas or liquid systems at finite Fermi degeneracies, 
which play a pivotal role in warm dense matter (WDM) \cite{Regan,Fletcher,Dharma1,Karasiev1} characterized by elevated temperatures and wide compression ranges.
A lot of astrophysical objects and materials under extreme experimental or environmental conditions are associated with the WDM, such as those highly compressed states observed in 
compact stars, planet cores, inertial confinement fusion, and laser ablation \cite{Ichimaru1}.
Theoretical descriptions by {\it ab initio} computer simulations for these materials are important both for interpreting experimental results and for obtaining insights 
into those parameter regions difficult to access experimentally, which 
could be performed in the framework of finite-temperature density functional theory (DFT) \cite{Dharma1,Karasiev1}.
In this formulation of finite-temperature DFT calculations, the accurate information on thermodynamics of electron fluid over a wide range of density and temperature is essential 
for constructing the input exchange-correlation potentials.

\par

Quantum Monte Carlo (QMC) simulations are in principle expected to give the exact solution to the correlational and thermodynamic quantities 
of electron gas \cite{Ceperley}.
As for the finite-temperature uniform electron gas, Brown {\it et al.} \cite{Brown1} first applied the restricted path integral Monte Carlo (RPIMC) method in coordinate space to the evaluation of 
correlational and thermodynamic functions, in which the fermion nodes of density matrix were fixed at those of ideal Fermi gas to avoid the sign problem \cite{Ceperley2}.
Brown {\it et al.} \cite{Brown2} also derived an analytic parametrization for the exchange-correlation free energy as a function of density and temperature of electron gas 
on the basis of their RPIMC data.
Unfortunately, their simulation results have been found to contain some systematic errors especially in the low-temperature and high-density regimes, and the parametrized expression 
for thermodynamic functions shows unphysical behaviors due to its irrelevant functional form.
Karasiev, Sjostrom, Dufty and Trickey 
(KSDT) \cite{Karasiev2} then proposed an improved fitting formula for the exchange-correlation free energy by performing appropriate interpolation 
with correct asymptotic limits.
In addition, other path integral Monte Carlo (PIMC) calculations, based on the permutation blocking path integral Monte Carlo (PB-PIMC) approach 
with a higher-order factorization of density matrix \cite{Dornheim,Groth} and 
the configuration path integral Monte Carlo (CPIMC) approach formulated in the Fock space of Slater determinants \cite{Groth,Schoof}, 
were also carried out to extend the validity of QMC calculations to wider parameter regions. 
Most recent advances in QMC simulations such as the density matrix quantum Monte Carlo (DMQMC) \cite{Malone} and the finite-size-corrected (FSC) PIMC \cite{Dornheim2} 
are ready to provide virtually exact results for the thermodynamic functions of finite-temperature electron gas over a wide range of density and temperature parameters. 
However, the difficulties associated with the fermion sign problem, the finite system size and time slice still remain, 
particularly in the low-temperature ($\theta \ll 1$) and high-density ($r_{s} \ll 1$) regimes,  
in spite of continuing efforts to overcome or correct them. 

\par

On the other hand, there has been a substantial progress since 1980s in the theoretical study of the correlational and thermodynamic properties of uniform electron gas or liquid 
at finite temperatures on the basis of analytical many-body theories \cite{Ichimaru1}.
The dielectric and thermodynamic functions have been calculated first in the random-phase approximation (RPA) \cite{Gupta1,Gupta2,Perrot2,Kanhere} and next in the approaches 
involving the local-field correction (LFC) \cite{Ichimaru1,Ichimaru2} which describes the strong-coupling effects beyond the RPA.
The latter includes those self-consistent integral equation approaches based on the parametrized LFC \cite{Dandrea}, the Singwi-Tosi-Land-Sj\"{o}lander (STLS) approximation 
\cite{Singwi,Tanaka2,Tanaka1}, the dynamical STLS approximation \cite{Schweng}, the Vashishta-Singwi (VS) approximation \cite{Vashishta,Sjostrom} and 
the modified convolution approximation (MCA) \cite{Tanaka3,Tago,Yan}.
Through satisfactory comparison with the existing Monte Carlo 
simulation data in the classical and ground-state limits, these analytic approaches have been expected to give fairly accurate predictions 
for the thermodynamic functions at any Fermi degeneracy of electrons, {\it e.g.}, within several $\%$ deviations in the exchange-correlation free energies from 
the ``true" values in the cases of STLS and MCA approximations \cite{Tanaka2,Tanaka1,Tanaka3}.
More recent hypernetted-chain (HNC) approach \cite{Tanaka4} provides one of the most reliable results for the thermodynamic functions, 
where reasonable agreements with the PIMC-based KSDT evaluations \cite{Karasiev2} are observed for the paramagnetic electron fluid (see Figs.\ 1 and 2 below).
However, it is remarkable that there are significant differences in the exchange-correlation free energy between the HNC and KSDT evaluations 
in the ferromagnetic state \cite{Tanaka4} (see Figs.\ 3 and 4 below), thus leaving a theoretical challenge.

\par

In earlier papers published in 1985-1986 \cite{Tanaka2,Tanaka1}, we derived an analytic expression for the equation of state (EOS) of the {\it paramagnetic} electron liquids at finite temperatures 
on the basis of the STLS approximation, which has been known to give the correlation energies in the ground state in good agreement with 
the Green's function Monte Carlo (GFMC) results \cite{Ceperley} at metallic densities ($2 \lesssim r_{s} \lesssim 6$).
Although it adequately described the thermodynamic properties of electron liquids over the whole region of the paramagnetic fluid phase, it had a room for further improvement 
with respect to the behaviors in the strong-coupling and the low-temperature regimes (see below).
Later, an improved expression for the thermodynamic functions in the paramagnetic state was proposed \cite{Ichimaru1,Ichimaru3} on the basis of the STLS interaction energies; 
since the appropriate corrections for the thermodynamic functions were performed in the strong-coupling and low-temperature regions, 
this improved EOS has been expected to provide very accurate evaluations for the exchange-correlation free energies and has actually proved to fairly agree with the PIMC results \cite{Sjostrom,Tanaka4}, 
while its detailed derivation has not been reported yet.

\par 

The present work firstly illustrates the derivation of the improved STLS-based EOS explicitly in the following Sec.\ II.
In addition to the derivation for the paramagnetic state, that for the ferromagnetic state is performed in an analogous manner, thus providing 
parametrized expressions for the interaction energy and the exchange-correlation free energy in the spin-polarized states.
Then, in Sec.\ III, the present STLS-based improved evaluation for the exchange-correlation free energies of finite-temperature electron liquids 
is compared with other existing evaluations in order to assess the current status for constructing the accurate EOS of the electron fluid 
over the whole regions of density and temperature at any Fermi degeneracy, Coulomb coupling and spin polarization.
Section IV concludes with some remarks.

\section{Derivation of equation of state}

The present approach begins with the derivation of the correlation functions for the uniform electron fluid at finite temperatures 
employing the dielectric formulation \cite{Ichimaru1,Ichimaru2}.
The wavenumber ($k$)-dependent static structure factor is calculated as 

\begin{equation}
S(k) = -\frac{k_{B}T}{n}\sum_{l=-\infty}^{\infty}\chi(k,z_{l}),
\end{equation}

\noindent
where the contributions from the discrete frequencies $z_{l}=2\pi ilk_{B}T/\hbar \enskip (l=0,\pm 1,\pm 2,\cdot\cdot\cdot)$ on the imaginary axis have been summed \cite{Tanaka1} 
for the complex-frequency-dependent, causal density response function \cite{Ichimaru1}, 

\begin{equation}
\chi(k,z) = \frac{\chi_{0}(k,z)}{1-v(k)\left[1-G(k,z)\right]\chi_{0}(k,z)}.
\end{equation}

\noindent
Here, $v(k)=4\pi e^{2}/k^{2}$ and the free-particle polarizability of electrons, 

\begin{equation}
\chi_{0}(k,z) = \sum_{\sigma}\sum_{\mbox{\boldmath $q$}}\frac{f_{\sigma}(\mbox{\boldmath $q$})-f_{\sigma}(\mbox{\boldmath $q$}+\mbox{\boldmath $k$})}
{\hbar z+\epsilon(\mbox{\boldmath $q$})-\epsilon(\mbox{\boldmath $q$}+\mbox{\boldmath $k$})+i0},
\end{equation}

\noindent
have been introduced 
with $\epsilon(q)=\hbar^{2}q^{2}/2m$; 
the Fermi distribution function for each spin species $\sigma$ (up or down) 

\begin{equation}
f_{\sigma}(q)=\left\{\exp\left[\frac{\epsilon(q)}{k_{B}T}-\alpha_{\sigma}\right]+1\right\}^{-1}
\end{equation}

\noindent
with the dimensionless chemical potential $\alpha_{\sigma}$ satisfies the normalization condition 

\begin{equation}
n_{\sigma} = \sum_{\mbox{\boldmath $k$}}f_{\sigma}(\mbox{\boldmath $k$})
\end{equation}

\noindent
with $n = \sum_{\sigma}n_{\sigma}$. 

\par

If we set $G(k,z) = 0$ in Eq.\ (7), we recover the random-phase approximation (RPA) expression for the density response function.
The local-field correction (LFC) $G(k,z)$ thus accounts for the strong-coupling effects beyond the RPA \cite{Ichimaru1,Ichimaru2} and plays an essential role 
to accurately describe the correlational and thermodynamic properties of electron liquids.

\par

In this study we rely on the static approximation to the LFC, i.e., $G(k,z) \simeq G(k)$, 
and employ the Singwi-Tosi-Land-Sj\"{o}lander (STLS) approximation \cite{Singwi} 
for a closure equation 
to determine $G(k)$ and $S(k)$ self-consistently in the wavenumber ($k$) space: 

\begin{equation}
G(k) = -\frac{1}{n}\sum_{\mbox{\boldmath $q$}}\frac{\mbox{\boldmath $k$}\cdot\mbox{\boldmath $q$}}{\mbox{\boldmath $q$}^{2}}
\left[S(\mbox{\boldmath $k$}-\mbox{\boldmath $q$})-1\right]. 
\end{equation}

\noindent
Given the static structure factor $S(k)$ obtained in this way, the interaction energy $E_{int}$ per electron is calculated as 

\begin{equation}
\frac{E_{int}}{N} = \frac{1}{2}\sum_{\mbox{\boldmath $k$}}v(k)\left[S(k)-1\right]. 
\end{equation}

\par

In the previous analysis reported in 1985-1986 \cite{Tanaka1,Tanaka2}, we computed the interaction energy $E_{int}$ of the paramagnetic finite-temperature electron liquids in the STLS approximation.
On the basis of the computed values for 70 combinations of the density and temperature parameters in the range of $r_{s} \leq 73.66$ and $\theta = 0.1$, 1 and 5, 
we constructed an analytic expression for $E_{int}$ as a function of $n$ and $T$, which reproduced the numerical results accurately and agreed with the known classical and ground-state limits satisfactorily.
However, this fitting expression also contained two difficulties regarding the behaviors in the strong-coupling regime associated with the inaccuracy in the STLS approximation and 
in the strong-degeneracy (low-temperature) regime associated with an inappropriate functional form to describe the Fermi-liquid parameters \cite{Tanaka3,Pines}.

\par

As is well known \cite{Ichimaru2,Tanaka3,Tanaka4}, the STLS scheme accurately predicted the values of the correlation energy obtained by the 
GFMC method \cite{Ceperley} for the degenerate electron system at metallic densities ($2 \lesssim r_{s} \lesssim 6$).
On the other hand, we also remark that the STLS scheme starts to show a tendency of underestimation of the magnitude of $E_{int}$ in comparison with 
the Monte Carlo (MC) simulation results \cite{Ceperley,Vosko,Slattery1,Slattery2} when the Coulomb coupling becomes strong.
These observations led us to reconstructing the EOS for the paramagnetic electron liquids \cite{Ichimaru1,Ichimaru3}, 
which will be reviewed in the following since its derivation procedure has not been reported ever.

\par

Let us define the deviation of the interaction energy computed in the STLS scheme from the ``exact" value as 

\begin{equation}
\delta = (\varepsilon - \varepsilon_{0})/\varepsilon_{0}.
\end{equation}

\noindent
Here $\varepsilon=-E_{int}/N(e^{2}/a)$ is the negative of reduced interaction energy per particle and $\varepsilon_{0}$ refers to that obtained in the STLS approximation.
When the MC value is employed for $\varepsilon$, for example, we find $\delta = 0.030$ \cite{Singwi,Tanaka3,Ceperley,Vosko} at $r_{s}=20$ in the ground state and 
$\delta = 0.079$ \cite{Yan,Slattery1,Slattery2} at $\Gamma = 50$ for the classical one-component plasma (OCP).

\par

To cope with such a systematic underestimation of the magnitude of the interaction energy in the STLS scheme, we attempted to evaluate $\delta$ as a function of 
$\theta$ and $\varepsilon_{0}$ in the intermediate to strong coupling regime.
We first parametrized $\delta(\varepsilon_{0},\theta)$ in the degenerate ($\theta \to 0$) and classical ($\theta \to \infty$) limits, where dependable MC results and 
their parametrizations \cite{Ceperley,Vosko,Slattery1,Slattery2} existed.
We thus found 

\begin{equation}
\delta(\varepsilon_{0},0) = 0.35466-2.32897\varepsilon_{0}+6.16040\varepsilon_{0}^{2}-7.96923\varepsilon_{0}^{3}+4.19679\varepsilon_{0}^{4} 
\end{equation}

\noindent
and 

\begin{equation}
\delta(\varepsilon_{0},\infty) = 1.71930-12.90154\varepsilon_{0}+35.28695\varepsilon_{0}^{2}-41.82031\varepsilon_{0}^{3}+18.33123\varepsilon_{0}^{4} 
\end{equation}

\noindent
for $0.45 \leq \varepsilon_{0} \leq 0.80$.
For $\varepsilon_{0} \leq 0.45$, where the Coulomb coupling is not so strong, we set $\delta(\varepsilon_{0},\theta)=0$.
For the regions of finite degrees of Fermi degeneracy, we then assumed such a hypothetical interpolation form as 

\begin{equation}
\delta(\varepsilon_{0},\theta) = \delta(\varepsilon_{0},0) + \left[\delta(\varepsilon_{0},\infty) - \delta(\varepsilon_{0},0)\right]\exp(-1/\theta).
\end{equation}

\par

An improved analytic expression for the interaction energy $\varepsilon_{int}=E_{int}/N(e^{2}/a)$ 
was then derived on the basis of the 70 STLS values so corrected according to Eqs.\ (13)-(16).
The result was 

\begin{equation}
\varepsilon_{int}(\Gamma,\theta,i) = -\frac{a_{i}(\theta)+b_{i}(\theta)\Gamma^{1/2}+c_{i}(\theta)\Gamma}{1+d_{i}(\theta)\Gamma^{1/2}+e_{i}(\theta)\Gamma},
\end{equation}

\noindent
where in the case of the paramagnetic state ($i = 0$), 

\begin{equation}
a_{0}(\theta) = \left(\frac{3}{2\pi}\right)^{2/3}\frac{0.75+3.04363\theta^{2}-0.092270\theta^{3}+1.70350\theta^{4}}
{1+8.31051\theta^{2}+5.1105\theta^{4}}\tanh\left(\frac{1}{\theta}\right),
\end{equation}

\begin{equation}
c_{0}(\theta) = \left[0.872496+0.025248\exp\left(-\frac{1}{\theta}\right)\right]e_{0}(\theta),
\end{equation}

\noindent
and a universal form, 

\begin{equation}
f(\theta) = F(\theta)\frac{x_{1}+x_{2}\theta^2+x_{3}\theta^4}{1+x_{4}\theta^2+x_{5}\theta^4}, 
\end{equation}

\noindent
is used for $b_{0}(\theta)$, $d_{0}(\theta)$ and $e_{0}(\theta)$ with the parametrizations given in Table 1.
This fitting formula reproduces the corrected 70 values of $\varepsilon_{int}$ with digressions of less than $0.6 \%$.

\par

In the weak coupling limit ($\Gamma \to 0$ or $r_{s} \to 0$), Eq.\ (17) approaches $-a_{0}(\theta)$, which represents the Hartree-Fock (exchange) contribution \cite{Perrot2}.
The coefficient $a_{0}(\theta)$ vanishes in the classical limit ($\theta \to \infty$) and the Debye-H\"{u}ckel term $-\sqrt{3}\Gamma^{1/2}/2$  then becomes 
the leading contribution.
In the strong coupling limit ($\Gamma \to \infty$ or $r_{s} \to \infty$), on the other hand, Eq.\ (17) approaches $-c_{0}(\theta)/e_{0}(\theta)$, which represents 
the Madelung-like contribution \cite{Ichimaru1,Ichimaru2}.
The ratio $c_{0}(\theta)/e_{0}(\theta)$ takes values of 0.872496 and 0.897744 in the degenerate and the classical limits, respectively, which are exactly the same as 
those found in the fitting formulas \cite{Vosko,Slattery1,Slattery2} for the MC data.
The formula (17) reproduces the MC values \cite{Slattery1,Slattery2} for $1 \leq \Gamma \leq 200$ with digressions of less than $0.5 \%$ in the classical limit.
In the degenerate (ground-state) limit, on the other hand, it agrees with the interaction energy derived from the GFMC fitting formula \cite{Vosko} 
for $r_{s} \leq 100$ within $0.4 \%$.

\par

The exchange-correlation free energy per particle in units of the Coulomb energy, $f_{xc}=F_{xc}/N(e^{2}/a)$, is then calculated by performing the coupling-constant ($\Gamma$) 
integration of $\varepsilon_{int}$ as 

\begin{equation}
f_{xc}(\Gamma,\theta) = \frac{1}{\Gamma}\int_{0}^{\Gamma}dx\ \varepsilon_{int}(x,\theta), 
\end{equation}

\noindent
thus leading to 

\begin{eqnarray}
f_{xc}(\Gamma,\theta,i) &=& -\frac{c_{i}}{e_{i}} - \frac{2}{e_{i}}\left(b_{i}-\frac{c_{i}d_{i}}{e_{i}}\right)\Gamma^{-1/2}  \nonumber \\
&-& \frac{1}{e_{i}\Gamma}\left[\left(a_{i}-\frac{c_{i}}{e_{i}}\right)-\frac{d_{i}}{e_{i}}\left(b_{i}-\frac{c_{i}d_{i}}{e_{i}}\right)\right] 
\ln\vert e_{i}\Gamma+d_{i}\Gamma^{1/2}+1\vert \nonumber \\
&+& \frac{2}{e_{i}(4e_{i}-d_{i}^2)^{1/2}\Gamma}\left[d_{i}\left(a_{i}-\frac{c_{i}}{e_{i}}\right)+\left(2-\frac{d_{i}^2}{e_{i}}\right)\left(b_{i}-\frac{c_{i}d_{i}}{e_{i}}\right)\right] \nonumber \\
 &\times& \left\{\tan^{-1}\left[\frac{2e_{i}\Gamma^{1/2}+d_{i}}{(4e_{i}-d_{i}^2)^{1/2}}\right]-\tan^{-1}\left[\frac{d_{i}}{(4e_{i}-d_{i}^2)^{1/2}}\right]\right\}
\end{eqnarray}

\noindent
for $i =$ 0 (paramagnetic state) and 1 (ferromagnetic state).

\par

In the first analytic expression \cite{Tanaka1,Tanaka2} for the EOS of the paramagnetic electron liquid, the reduced exchange-correlation free energy $f_{xc}$ 
had such a form as expanded in terms of the power of $\theta^{1/2}$ in the low-temperature limit ($\theta \to 0$).
This fact causes some difficulties in the calculations of those thermodynamic quantities derived through differentiations of the free energy with respect to 
the temperature, such as the entropy and the specific heat.
We have therefore adopted in the expressions above those functional forms which have a low-temperature expansion beginning with the terms of order $\theta^{2}$, 
thus corresponding with the consequence of the Fermi-liquid theory \cite{Tanaka3,Pines}.

\par

Next, we proceed to the derivation of the EOS for the fully spin-polarized (ferromagnetic) state with $\zeta=1$.
The present work has newly computed the interaction energy $E_{int}$ of the ferromagnetic finite-temperature electron liquids in the STLS approximation for 
70 combinations of the density and temperature parameters in the range of $r_{s} \leq 73.66$ and $\theta = 0.1$, 1 and 5, which are identical combinations to those for the paramagnetic state.
The calculated values for the STLS interaction energies are compiled in Supplemental Material \cite{SM}.
The following procedure for correcting the STLS interaction energies has been carried out in an analogous way to the paramagnetic case.
First, a fitting expression for the difference of the interaction energies between the STLS and GFMC \cite{Vosko} values has been found as 

\begin{equation}
\delta(\varepsilon_{0},0) = 2.76114-21.32500\varepsilon_{0}+60.93707\varepsilon_{0}^{2}-76.26419\varepsilon_{0}^{3}+35.23686\varepsilon_{0}^{4} 
\end{equation}

\noindent
for $0.45 \leq \varepsilon_{0} \leq 0.80$ in the ground state.
Then, using Eqs.\ (13), (16), (15) and (23), we obtain the corrected 70 values for the interaction energies in the ferromagnetic state.
Analytically fitting these corrected values, one can derive an expression for $\varepsilon_{int}$ as Eq.\ (17), where $a_{i}(\theta) - e_{i}(\theta)$ 
in the case of the ferromagnetic state ($i=1$) are given by 

\begin{equation}
a_{1}(\theta) = 1.25992 a_{0}(0.629961\theta),
\end{equation}

\begin{equation}
c_{1}(\theta) = \left[0.893638+0.004106\exp\left(-\frac{1}{\theta}\right)\right]e_{1}(\theta),
\end{equation}

\noindent
and Eq.\ (20) for $b_{1}(\theta)$, $d_{1}(\theta)$ and $e_{1}(\theta)$ with the parameters compiled in Table 1.
This fitting formula reproduces the corrected 70 values of $\varepsilon_{int}$ with the maximum and average digressions of $1.38 \%$ and $0.25 \%$, respectively.

\par

In the weak coupling limit ($\Gamma \to 0$ or $r_{s} \to 0$), Eq.\ (17) for the ferromagnetic state approaches $-a_{1}(\theta)$, which represents the Hartree-Fock (exchange) contribution \cite{Perrot2}.
The coefficient $a_{1}(\theta)$ vanishes in the classical limit ($\theta \to \infty$) and the Debye-H\"{u}ckel term $-\sqrt{3}\Gamma^{1/2}/2$  then becomes 
the leading contribution.
In the strong coupling limit ($\Gamma \to \infty$ or $r_{s} \to \infty$), on the other hand, Eq.\ (17) approaches $-c_{1}(\theta)/e_{1}(\theta)$, which 
takes values of $-0.893638$ and $-0.897744$ in the degenerate and the classical limits, respectively, the same as 
those found in the fitting formulas \cite{Vosko,Slattery1,Slattery2} for the MC data.
The formula (17) reproduces the MC values \cite{Slattery1,Slattery2} for $1 \leq \Gamma \leq 200$ with digressions of less than $0.9 \%$ in the classical limit.
In the degenerate (ground-state) limit, on the other hand, it agrees with the interaction energy derived from the GFMC fitting formula \cite{Vosko} 
for $r_{s} \leq 100$ within $0.5 \%$.
The exchange-correlation free energy per particle in units of the Coulomb energy, $f_{xc}=F_{xc}/N(e^{2}/a)$, is then given by Eq.\ (22).

\par

We have thus derived the analytic expressions for $\varepsilon_{int}$ and $f_{xc}$ as the functions of $\theta$ and $\Gamma$ (or $r_{s}$) both in the paramagnetic ($\zeta = 0$) and ferromagnetic ($\zeta = 1$) states. 
In order to obtain the expressions available at any spin polarization ($0 \leq \zeta \leq 1$), we can use the interpolation scheme developed in the MCA study \cite{Tanaka3} 
in which the calculations for $\varepsilon_{int}$ at $\zeta =$ 0, 0.2, 0.5, 0.8 and 1 were carried out. 
It is expected that the MCA interpolation function which depends on $\zeta$, $r_{s}$ and $\theta$ would work well also in the present case, 
called the improved STLS (iSTLS) scheme hereafter, because the differences in $\varepsilon_{int}$ and $f_{xc}$ 
between the iSTLS and MCA approximations are very minor over the whole $r_{s}-\theta$ regions and the functional dependence on the degree of spin polarization is 
expected to be smooth and simple. 
In this scheme, the exchange-correlation free energy at any spin polarization $\zeta$ is expressed by 

\begin{equation}
f_{xc}(\Gamma,\theta,\zeta) = (1-\zeta^6)f_{xc}(\Gamma,\theta,0)+\zeta^{6}f_{xc}(\Gamma,\theta,1)+
\left(\frac{1}{2}\zeta^{2}+\frac{5}{108}\zeta^{4}-\frac{59}{108}\zeta^{6}\right)\frac{\alpha_{xc}}{N(e^{2}/a)}.
\end{equation}

\noindent
The parametrized form for $\alpha_{xc}/N(e^{2}/a)$ as a function of $\theta$ and $\Gamma$ (or $r_{s}$) has been given in Ref. \cite{Tanaka3}.

\section{Comparisons with other evaluations}

Figure 1 (a)-(c) illustrate the various evaluations for the reduced exchange-correlation free energy $f_{xc}$ in the paramegnetic state as functions of $r_{s}$ at $\theta =$ 0.5, 1 and 4, where 
the iSTLS evaluation \cite{Ichimaru1} is compared with the results by the HNC \cite{Tanaka4}, MCA \cite{Tanaka3}, KSDT \cite{Karasiev2} and finite-size-corrected (FSC) QMC \cite{Dornheim2} schemes.
As seen in Fig.\ 1(b) for $\theta = 1$, all these evaluations give virtually identical results at intermediate Fermi degeneracy, thus directing us to a right answer.
At $\theta = 0.5$ in Fig.\ 1(a), the HNC and MCA values show slightly downward deviations from others by a few $\%$ for $1 \lesssim r_{s} \lesssim 10$, 
which may be due to an overestimation of the exchange-correlation hole at short ranges \cite{Tanaka4}.
At $\theta = 4$ in Fig.\ 1(c), on the other hand, we observe that the KSDT values show a downward deviation from others for $0.01 \lesssim r_{s} \lesssim 1$.
The FSC-QMC parametrization \cite{Dornheim2} which is based on the nearly exact PIMC simulations for $0.5 \leq \theta \leq 8$ and $0.1 \leq r_{s} \leq 10$ may be regarded as 
the most dependable standard among these evaluations for $f_{xc}$.
One can then compare the iSTLS and KSDT parametrizations for $f_{xc}$ to the FSC-QMC values for all the 45 cases compiled in the latter study \cite{Dornheim2}.
The mean absolute relative error (MARE) in the iSTLS evaluation (which was published in 1987 \cite{Ichimaru1}) is $1.4 \%$, which is superior to the MARE in the KSDT evaluation, $2.7 \%$, 
derived on the basis of the RPIMC calculations \cite{Brown1,Brown2}.

\par

Figure 2 (a)-(c) illustrate the various evaluations for $f_{xc}$ in the paramegnetic state as functions of $\theta$ at $r_{s} =$ 1, 6 and 10, respectively.
In addition to the iSTLS, HNC, MCA, KSDT and FSC-QMC values, the evaluations by the (original) STLS \cite{Tanaka1,Tanaka2}, Vashishta-Singwi (VS) \cite{Sjostrom}, and 
classical-map hypernetted-chain (CHNC) \cite{Dharma2,Perrot1} schemes are also depicted in these figures.
Overall, the VS and CHNC evaluations show slightly downward and upward digressions from others, respectively, which has been remarked in an earlier study \cite{Tanaka4}.
The original STLS parametrization \cite{Tanaka1,Tanaka2} with an inappropriate functional form regarding the low-$\theta$ dependence shows a spurious decrease in $f_{xc}$ 
for the low-temperature range of $0.1 \lesssim \theta \lesssim 1$, which has been corrected in the iSTLS parametrization \cite{Ichimaru1} as addressed above.
Besides, slight deviations of the HNC and MCA values from the iSTLS, KSDT and FSC-QMC values for $0.1 \lesssim \theta \lesssim 1$ are analogous to those observed in Fig.\ 1(a).

\par

As for the ferromagnetic state, Fig.\ 3 (a)-(c) show the evaluations of $f_{xc}$ as functions of $r_{s}$ at $\theta =$ 0.4, 1 and 4, respectively, where 
the present iSTLS values based on Eq.\ (22) are compared with the HNC \cite{Tanaka4}, MCA \cite{Tanaka3}, KSDT \cite{Karasiev2} and CHNC \cite{Dharma2,Perrot1} values.
Slight deviations of the HNC and MCA values from the iSTLS evaluation for $r_{s} \gtrsim 1$ at $\theta = 0.4$ in Fig.\ 3(a) are analogous to those observed for the paramagnetic state in Figs.\ 1 and 2.
A significant feature in Fig.\ 3 (a)-(c) is the digression of the KSDT values from the iSTLS, HNC and MCA values, {\it i.e.}, downward and upward deviations 
for $0.01 \lesssim r_{s} \lesssim 1$ and $1 \lesssim r_{s} \lesssim 100$, respectively.
In the case of Fig.\ 3(b) for $\theta = 1$, the iSTLS values lie between the HNC/MCA and KSDT values for $1 \lesssim r_{s} \lesssim 100$, but more closely to the former; 
{\it e.g.}, the MCA, HNC and iSTLS values deviate from the KSDT value by 6.5, 5.5 and $4.2 \%$, respectively, at $r_{s} = 10$.
We would thus like to suggest the necessity of re-examination on the EOS by the QMC methods for the ferromagnetic state \cite{Tanaka4}.

\par

The $\theta$ dependence of $f_{xc}$ in the ferromagnetic state is compared among the various schemes at $r_{s} =$ 1, 6 and 10 in Fig.\ 4 (a)-(c), respectively.
These figures again indicate that the agreement among various approaches is unsatisfactory compared to the paramagnetic case. 
It is noted that in the low-temperature region ($\theta \ll 1$) the iSTLS evaluation is more akin to the KSDT evaluation than to the HNC and MCA values.
Finally, Fig.\ 5 illustrates the spin polarization ($\zeta$) dependences of $f_{xc}$ at $\theta = 1$ and $r_{s} = 6$ calculated by the iSTLS, KSDT \cite{Karasiev2} 
and CHNC \cite{Dharma2,Perrot1} parametrizations.
We observe that the iSTLS evaluation on the basis of the MCA parametrization 
for $\zeta$ dependence \cite{Tanaka3}, Eq.\ (26), shows a smooth interpolation between the paramegnetic and ferromagnetic states 
as well as the KSDT and CHNC parametrizations.
Figure 5 also indicates how both the evaluations of iSTLS and KSDT deviate from each other more significantly as the 
degree of spin polarization proceeds toward higher values.

\section{Conclusions}

The present work has been focused on a construction of accurate expression for exchange-correlation free energy $f_{xc}$ of homogeneous electron fluid at any combination of 
Coulomb coupling, Fermi degeneracy and spin polarization.
The approach is based on the STLS integral-equation method in the dielectric formulation that is known to be accurate in the weak to intermediate coupling regime, and 
an appropriate strong-coupling correction has been incorporated into the proposed fitting expression for the EOS. 
The analytic formula for $f_{xc}$ is identical to that proposed previously \cite{Ichimaru1,Ichimaru3} in the paramagnetic state, but brand-new for the ferromagnetic state, 
whose derivations are explicitly illustrated for the first time in this work.
Its extension to any degree of spin polarization ($0 < \zeta < 1$) is then performed with the aid of interpolation formula studied in the MCA scheme \cite{Tanaka3}.
The proposed (iSTLS) EOS is thus expected to be accurate over the whole parameter regions of density, temperature and spin polarization of electron fluid, 
which would provide useful and reliable inputs to the finite-temperature DFT calculations.
While the present iSTLS expression has been derived through an analytic approach with a strong-coupling correction owing to the MC data in the classical and degenerate limits, 
it compares fairly well with the PIMC data at finite Fermi degeneracies \cite{Karasiev2,Dornheim2} in the paramagnetic state when the latter evaluations are available.
Since the accurate QMC data are currently missing for $\theta \lesssim 0.5$ in the thermodynamic limit of the paramagnetic electron fluid due to the fermion sign problem, 
the iSTLS data would provide a relevant guideline for finding accurate thermodynamic functions in these low-temperature regimes. 
On the other hand, the consensus among available analytic and simulation results is unsatisfactory in the case of the ferromagnetic state, 
as seen in Figs.\ 3 and 4 above, which manifests a challenge to QMC simulators.

\section*{Acknowledgements}

\noindent
The author would like to acknowledge the Grants-in-Aid for Scientific Research (No.\ 26460035) from the Ministry of Education, Cultute, Sports, Science and Technology (MEXT).
The numerical computations in the present work were carried out by the IBM eServer p7 model 755 at the Information Science and Technology Center of Kobe University.

\section*{Supplemental Material}

\noindent
The interaction energies of spin-polarized ($\zeta = 1$) electron liquids at $\theta =$ 0.1, 1 and 5 calculated with the STLS approximation 
are listed in Supplemental Material.

\newpage

\section*{Table}

\noindent
Table 1.\ Values of $x_{j}$  ($j = 1 - 5$) and functional forms of $F(\theta)$ in Eq.\ (20) for the coefficients $b_{i}$, $d_{i}$ and $e_{i}$ ($i = 0, 1$) appearing in Eqs.\ (17) and (22). 

\begin{center}
\begin{tabular}{ccccccc} \hline
$f(\theta)$ & $F(\theta)$ & $x_{1}$ & $x_{2}$ & $x_{3}$ & $x_{4}$ & $x_{5}$ \\ \hline
$b_{0}(\theta)$ & $\sqrt{\theta}\tanh(1/\sqrt{\theta})$ & 0.341308 & 12.070873 & 1.148889 & 10.495346 & 1.326623 \\ 
$d_{0}(\theta)$ & $\sqrt{\theta}\tanh(1/\sqrt{\theta})$ & 0.614925 & 16.996055 & 1.489056 & 10.109350 & 1.221840 \\
$e_{0}(\theta)$ & $\theta\tanh(1/\theta)$ & 0.539409 & 2.522206 & 0.178484 & 2.555501 & 0.146319 \\
$b_{1}(\theta)$ & $\sqrt{\theta}\tanh(1/\sqrt{\theta})$ & 0.432217 & 12.364621 & 0.850624 & 10.231916 & 0.982216 \\ 
$d_{1}(\theta)$ & $\sqrt{\theta}\tanh(1/\sqrt{\theta})$ & 0.680892 & 17.208524 & 1.562826 & 9.892598 & 1.124675 \\
$e_{1}(\theta)$ & $\theta\tanh(1/\theta)$ & 0.309934 & 2.251028 & 0.216962 & 2.736885 & 0.134352 \\ \hline
\end{tabular}
\end{center}

\newpage


\section*{Figure Captions}


\begin{figure}[h]
\caption{Exchange-correlation free energies per electron in units of $e^2/a$, $f_{xc}$, as functions of $r_{s}$ at fixed $\theta$ in the paramagnetic state ($\zeta = 0$). 
The evaluations by the iSTLS \cite{Ichimaru1,Ichimaru3}, HNC \cite{Tanaka4}, MCA \cite{Tanaka3}, KSDT \cite{Karasiev2} and finite-size-corrected (FSC) QMC \cite{Dornheim2} 
schemes are represented by blue solid line, magenta dashed line, orange dotted line, 
green dashed line and red solid line, respectively. 
(a) $\theta = 0.5$. 
(b) $\theta = 1$. 
(c) $\theta = 4$. 
}

\caption{Exchange-correlation free energies per electron in units of $e^2/a$, $f_{xc}$, as functions of $\theta$ at fixed $r_{s}$ in the paramagnetic state ($\zeta = 0$). 
The evaluations by the iSTLS \cite{Ichimaru1,Ichimaru3}, HNC \cite{Tanaka4}, MCA \cite{Tanaka3}, KSDT \cite{Karasiev2}, STLS \cite{Tanaka1,Tanaka2}, Vashishta-Singwi (VS) \cite{Sjostrom}, 
classical-map HNC (CHNC) \cite{Dharma2,Perrot1} and finite-size-corrected (FSC) QMC \cite{Dornheim2} schemes 
are represented by blue solid line, magenta dashed line, orange dotted line, green dashed line, grey solid line, yellow-green dotted line, 
cyan filled circles and red filled circles, respectively. 
(a) $r_{s} = 1$. 
(b) $r_{s} = 6$. 
(c) $r_{s} = 10$. 
}

\caption{Exchange-correlation free energies $f_{xc}$ as functions of $r_{s}$ at fixed $\theta$ in the ferromagnetic state ($\zeta = 1$). 
The evaluations by the iSTLS, HNC \cite{Tanaka4}, MCA \cite{Tanaka3}, KSDT \cite{Karasiev2} and CHNC \cite{Dharma2,Perrot1} 
schemes are represented by blue solid line, magenta dashed line, orange dotted line, 
green dashed line and cyan filled circles, respectively. 
(a) $\theta = 0.4$. 
(b) $\theta = 1$. 
(c) $\theta = 4$. 
It is noted that ``$\theta =1$" in the present study corresponds to $t=2^{-2/3}=0.629961$ in the KSDT fitting \cite{Karasiev2} for the ferromagnetic state. 
}

\caption{Exchange-correlation free energies $f_{xc}$ as functions of $\theta$ at fixed $r_{s}$ in the ferromagnetic state ($\zeta = 1$). 
The evaluations by the iSTLS, HNC \cite{Tanaka4}, MCA \cite{Tanaka3}, KSDT \cite{Karasiev2} and 
CHNC \cite{Dharma2,Perrot1} schemes are represented by blue solid line, magenta dashed line, orange dotted line, green dashed line 
and cyan filled circles, respectively. 
(a) $r_{s} = 1$. 
(b) $r_{s} = 6$. 
(c) $r_{s} = 10$. 
}

\caption{Spin polarization ($\zeta$) dependence of $f_{xc}$ at $\theta = 1$ and $r_{s} = 6$ evaluated by 
the iSTLS (blue solid line), KSDT (green dashed line) \cite{Karasiev2} and CHNC (cyan dotted line) \cite{Dharma2,Perrot1} parametrizations.
} 
\end{figure}

\end{document}